\documentclass[aps,prd,twocolumn,superscriptaddress,nofootinbib,10pt]{revtex4-2}

\usepackage{graphicx} 
\usepackage{dcolumn} 
\usepackage{bm} 
\usepackage{slashed}
\usepackage{xcolor}
\usepackage{colortbl}
\definecolor{CBPurple}{RGB}{34,139,34}
\usepackage{amssymb}
\usepackage{xspace}
\usepackage{pifont}
\newcommand{\cmark}{\ding{51}}%
\newcommand{\xmark}{\ding{55}}%

\usepackage{comment}
\usepackage{amsmath}
\usepackage{lmodern}
\usepackage{dsfont}
\usepackage{hyperref}
 \usepackage[capitalise]{cleveref}
\usepackage{orcidlink}
\usepackage{makecell}
\usepackage{subcaption}
\usepackage{multirow}
\usepackage{booktabs}
\usepackage{siunitx}

\usepackage[letterpaper,top=1.0in, inner=0.5in, outer=0.5in,bottom=1.0in,,headheight=0pt,headsep=20pt,centering]{geometry}

\def\ee{$e^+e^-$}

\newcommand*{\pnet}{\ensuremath{\textsc{ParticleNet}}}
\newcommand*{\pnetee}{\ensuremath{\textsc{ParticleNetIdea}}}
\newcommand*{\ParT}{\ensuremath{\textsc{ParT}}}

\newcommand*{\py}{\ensuremath{\textsc{Pythia6}}}
\newcommand*{\whiz}{\ensuremath{\textsc{Whizard-3.1.4}}}

\newcommand*{\delphes}{\ensuremath{\textsc{Delphes}}}
\newcommand*{\geant}{\ensuremath{\textsc{Geant4}}}

\newcommand*{\fastjet}{\ensuremath{\textsc{FastJet-3.3.4}}}

\newcommand*{\weaver}{\ensuremath{\textsc{weaver}}}
\newcommand*{\weavercore}{\ensuremath{\textsc{weaver-core}}}
\newcommand*{\ranger}{\ensuremath{\textsc{ranger}}}

\newcommand*{\trackcovMod}{\ensuremath{\texttt{TrackCovariance}}}
\newcommand*{\tofMod}{\ensuremath{\texttt{TimeOfFlight}}}
\newcommand*{\ccMod}{\ensuremath{\texttt{ClusterCounting}}}

\def\TeV{\ifmmode {\mathrm{\ Te\kern -0.1em V}}\else
                   \textrm{Te\kern -0.1em V}\fi}%
\def\GeV{\ifmmode {\mathrm{\ Ge\kern -0.1em V}}\else
                   \textrm{Ge\kern -0.1em V}\fi}%
\def\MeV{\ifmmode {\mathrm{\ Me\kern -0.1em V}}\else
                   \textrm{Me\kern -0.1em V}\fi}%
\def\keV{\ifmmode {\mathrm{\ ke\kern -0.1em V}}\else
                   \textrm{ke\kern -0.1em V}\fi}%
\def\eV{\ifmmode  {\mathrm{\ e\kern -0.1em V}}\else
                   \textrm{e\kern -0.1em V}\fi}%

\definecolor{CBBlue}{RGB}{0,114,178}       
\definecolor{CBOrange}{RGB}{230,159,0}     
\definecolor{CBPurple}{RGB}{204,121,167}   
\begin{document}

\title{Evaluating the Impact of Detector Design on Jet Flavor Tagging for Future Colliders}

\author{Dimitrios Ntounis\orcidlink{0009-0008-1063-5620}}  
\email{dntounis@slac.stanford.edu}

\affiliation{SLAC National Accelerator Laboratory, 2575 Sand Hill Road, Menlo Park, California 94025, USA  \\ Stanford University, 450 Jane Stanford Way, Stanford, California 94305, USA}

\author{Loukas Gouskos\,\orcidlink{0000-0002-9547-7471}}
\email{loukas\_gouskos@brown.edu}

\affiliation{Department of Physics, Brown University, Providence, Rhode Island 02912, USA}

\author{Caterina Vernieri\,\orcidlink{0000-0002-0235-1053}}
\email{caterina@slac.stanford.edu}

\affiliation{SLAC National Accelerator Laboratory, 2575 Sand Hill Road, Menlo Park, California 94025, USA  \\ Stanford University, 450 Jane Stanford Way, Stanford, California 94305, USA}


\begin{abstract}
Jet flavor tagging is of utmost importance for unlocking the full physics potential of any future collider experiment. The performance of any jet flavor identification algorithm depends both on its underlying architecture and on the detector's design and capabilities. In this work, we present an analysis of the dependence of jet tagging algorithm performance on three detector designs being considered for future $e^{+}e^{-}$ colliders. To fully exploit the potential of these detector concepts, we utilize a graph neural network-based jet tagging algorithm. In addition, we evaluate the impact on the jet tagging performance of variations in the tracking and calorimeter systems for one of these detector concepts, the SiD detector, as well as the dependence on the center-of-mass energy.
\end{abstract}

\maketitle

\section{Introduction}
\label{sec:intro}

The primary goal of all proposed future \ee \ colliders is the measurement of Standard Model (SM) parameters with very high precision. For many parameters, such as the coupling of the Higgs boson to the bottom ($b$), charm ($c$) and strange ($s$) quarks, as well as gluons ($g$), the ability to identify the flavor of the parton that initiated each jet, known as jet flavor tagging, is of crucial importance~\cite{Thomson:2015jda,ILC_Snowmass,FCC:2018evy,Azzi:2021gwg}.

Jet flavor tagging algorithms aim to exploit differences in the experimental signature of jets, coming from the fragmentation and hadronization patterns of the initiating parton, in order to accurately assign jet flavors. In particular, gluon-initiated jets tend to contain more particles than light-quark  $(q)  $ jets\footnote{In the following, we collectively call $u$- and $d$-quark initiated jets as $q$-jets or light-quark jets.}, have more uniform energy fragmentation and be less collimated. Heavy flavor jets originating from $b$ ($c$) quarks contain $B$ ($D$) hadrons with typical lifetimes of $\sim 1.5 \  (0.4-1.0) \ \mathrm{ps}$~\cite{HeavyFlavorAveragingGroupHFLAV:2024ctg} that travel measurable distances in the detector before decaying, leading to displaced tracks and secondary vertices. Finally, $s$-quarks, which lie somewhere between $q$-jets and heavy flavor jets in terms of their experimental footprint, tend to show a larger fraction of their energy carried by strange hadrons, mostly kaons~\cite{Albert:2022mpk}. The ability to identify jets originating from strange quarks is particularly important for measuring the $H \rightarrow s\bar{s}$ decay at future \ee \ colliders, providing direct access to the Higgs coupling to strange quarks.

Modern advances in machine learning have enabled a new generation of advanced jet flavor taggers~\cite{Bols:2020bkb,Qu:2019gqs,Qu:2022mxj,ATLAS:2022rkn}, which combine low-level information (i.e., jet constituents), typically in the form of reconstructed Particle-Flow Objects (PFOs)~\cite{CMS:2017yfk,Thomson:2009rp}, through deep neural network architectures in order to fully exploit discriminating characteristics such as the above. This low-level information comprises track-related properties and Particle Identification (PID) of the jet constituents, with each sub-detector system contributing differently to each. For instance,  low-material-budget, high-granularity vertex and tracking detectors are crucial for heavy flavor tagging, which requires efficient and precise track and vertex reconstruction. Additionally, powerful PID capabilities envisaged for some detector concepts, through ionization energy loss and/or time-of-flight (TOF) measurements, are expected to provide additional discriminating power between different jet species. This is particularly relevant for $s$-tagging in $H\rightarrow s\bar{s}$ decays, where the ability to distinguish charged kaons from protons and pions at momenta up to tens of GeV is crucial, as kaons in $s-$jets typically carry a larger fraction of the jet momentum compared to $q$-jets.

The main focus of this work is the study of the dependence of jet flavor tagging on the underlying detector design. While previous studies have examined specific aspects of flavor tagging, particularly vertex detector geometry, and assuming a specific detector choice, this work provides a unified analysis framework that enables direct comparisons between detector concepts.  Using the \delphes~\cite{deFavereau:2013fsa} fast simulation framework and the \pnet \  jet flavor tagger~\cite{Qu:2019gqs}, we evaluate the jet flavor discriminating power for three baseline detector concepts for future \ee \ colliders in a consistent manner, ensuring a fair comparison through identical samples, tagging algorithm and analysis methods. Beyond comparing baseline designs, we also explore the impact of variations of several detector parameters, namely the vertex detector geometry and calorimetry resolution and segmentation. This flexible framework enables rapid re-evaluation of flavor tagging performance as detector designs evolve, providing valuable feedback for detector optimization studies. The rest of the paper is organized as follows: an overview of the simulation and flavor tagging framework is presented in ~\cref{sec:framework}. The results of our analysis are presented and discussed in~\cref{sec:results}. Finally,~\cref{sec:conclusions} contains our conclusive remarks, together with perspectives for future work.

\section{Analysis Framework}
\label{sec:framework}
\subsection{\delphes \  Fast Simulation}

The detector response was simulated using the \delphes \  fast simulation framework. In \delphes, parameterized detector response formulas, such as those for the tracking efficiency of charged particles and the single-particle energy resolution of the calorimeters, typically derived from more detailed full-detector simulations, are used to statistically smear hard-scatter-level quantities. Although this may lead to more optimistic performance in physics benchmarks, it provides a modular, resource-efficient, and significantly faster framework, relevant for the scope of this study. While such fast simulation results provide valuable initial insights into detector optimization, future studies using full, \geant~\cite{GEANT4:2002zbu}  based, simulation, should be pursued to validate such findings.

In this study, we also take advantage of the latest developments in \delphes, such as the \tofMod~\cite{tof_module},  \ccMod~\cite{cc_module}, and \trackcovMod~\cite{tc_module} modules. The former two provide PID information by inferring particle masses through TOF measurement or ionization cluster counting, while the latter provides an estimate for the track parameters, and associated covariance matrix, for each charged particle, given a geometry description of the detector's tracking system. \trackcovMod \  is of particular importance for the development of a jet flavor tagger, as track-related variables have been shown to bear the largest discriminating power~\cite{Blekman:2024wyf}, due to the majority of the jet's energy (roughly $60\%$~\cite{Marshall:2013bda}) being carried by charged particles. Technical details on these \delphes \ modules are provided in~\cite{Bedeschi:2022rnj}.

\subsection{Detector Models}
\label{subsec:detector_models}

In this work, we evaluate jet flavor tagging performance for two detectors envisaged for future $e^{+}e^{-}$ colliders: SiD~\cite{Aihara:2009ad} and IDEA~\cite{FCC:2018evy}, designed for the International Linear Collider (ILC) and Future Circular Collider (FCC-ee) respectively. SiD is a compact detector based on a powerful and robust silicon vertex and tracker, and highly segmented ECAL and HCAL, all contained within a strong 5T solenoid field. IDEA, on the other hand, relies on a tracking system comprised of a silicon innermost vertex detector and a gaseous drift chamber surrounded by a silicon wrapper, all contained within a 2T solenoid. A dual-readout calorimeter is expected to provide excellent electromagnetic and hadronic energy resolution. 

 These detectors were selected with the purpose of comparing the physics performance of detectors across different collider concepts and, more importantly, to evaluate the impact of PID capabilities on jet flavor tagging performance: while IDEA features dedicated particle identification through TOF and cluster counting $(\mathrm{d}N/\mathrm{d}x)$ measurements, SiD relies on tracking and calorimetry information.

For IDEA, the latest official \delphes \ model was used, as found in~\cite{delphes_card_idea}. An additional detector model, named FCCeeDetWithSiTracking~\cite{delphes_card_FCCeeDetWithSiTracking}, is also available, corresponding to an IDEA-like detector with its tracking system replaced with that of another concept for FCC-ee: the CLD~\cite{Bacchetta:2019fmz}. Both of these detector models are used to study the impact of the two different tracking system choices on flavor tagging.

For SiD, the existing official \delphes \ model available~\cite{Potter:2016pgp} did not include an implementation of the latest \delphes \ modules introduced earlier. For that reason, a new SiD \delphes \  configuration was generated, with the SiD vertex and tracker geometry, as specified in the Detailed Baseline Design (DBD)~\cite{Behnke:2013lya}, implemented in \trackcovMod. Modifications with respect to the design parameters in~\cite{Behnke:2013lya} were made in the following aspects in order to account for foreseen improvements due to technological advances since the DBD, as laid out in~\cite{Breidenbach:2021sdo}: (a) the single-hit spatial resolution in the vertex (tracker) system was assumed to be $3 \  (7)$ \textmu m, matching the target for recent Monolithic Active Pixel sensors (MAPS) research and development efforts~\cite{Apadula:2022ees}, (b) the single-particle energy resolution and spatial resolution of the electromagnetic calorimeter (ECAL) were adjusted to account for potential gains from a large-area MAPS digital ECAL~\cite{Brau:2022sxr,Brau:2024}, and (c) the single-particle energy resolution and spatial resolution of the hadronic calorimeter (HCAL) were adjusted based on the latest results by the CALICE collaboration~\cite{Sefkow:2015hna} for a prototype with scintillator tiles read out by silicon photomultipliers. The SiD \delphes \  model developed for the purposes of this study can be found in~\cite{delphes_card_sid}. A summary of key specifications of the SiD, IDEA and FCCeeDetWithSiTracking detectors, as implemented in \delphes, is given in Table~\ref{tab:detector_specs}.

\begin{table*}[htbp]
    \centering
    \renewcommand{\arraystretch}{1.1}
      \caption[]{Overview of key specifications  for the SiD, IDEA and {FCCeeDetWithSiTracking} detectors, as implemented in the corresponding \delphes \  models. }
    \label{tab:detector_specs}
    \hspace*{-0.9cm}
\begin{tabular}{l c  c c c}
\toprule
\textbf{Parameter} & \textbf{Symbol [unit]}  & \textbf{SiD} & \textbf{IDEA} & \textbf{FCCeeDetWithSiTracking}\\ 
\midrule 
 Magnetic field & $B$ [T]  & 5 & \multicolumn{2}{c}{2}  \\
 TOF Resolution & $\sigma_{\mathrm{TOF}}$ [ps]  & $10^{3}$ & \multicolumn{2}{c}{30}  \\
Cluster Counting & $\mathrm{d}N/\mathrm{d}x$  & \xmark & \multicolumn{2}{c}{\cmark}  \\
\midrule 
\multicolumn{5}{c}{\emph{Vertex Barrel}} \\ 
\midrule
Number of layers & $N_{\mathrm{layers}}$ & 5 & 5 & 3 \\
Radial range from IP\footnote{Interaction Point} & r[cm] & 1.4-18.1 & 1.37-31.5 &  1.2-3.15 \\
Single-hit spatial resolution & $\sigma_{xy}$ [\textmu m$^2$] & $3 \times 3$ &  \begin{tabular}[c]{@{}c@{}} $3 \times 3$ (first 3 layers)\\ $7 \times 7$ (last 2 layers)\end{tabular} & $3 \times 3$  \\
\midrule 
\multicolumn{5}{c}{\emph{Tracker Barrel}} \\ 
\midrule
Number of layers & $N_{\mathrm{layers}}$ & 5 &  \begin{tabular}[c]{@{}c@{}} $ 112$ (Drift Chamber)\\ $2$ (Si wrapper)\end{tabular}  & 6 \\
Radial range & r[cm] & 21.95-122.0 & 34.5-245.0  &  12.7-212.6 \\
Single-hit spatial resolution & $\sigma_{xy}$ [\textmu m$^2$] & $7 \times 7$ & \begin{tabular}[c]{@{}c@{}} $100 \times 100$ (Drift Chamber)\\ $7 \times 90$ (Si wrapper)\end{tabular}   & $7 \times 90$  \\

\midrule  
\multicolumn{5}{c}{\emph{ECAL}} \\ 
\midrule 

Transverse Spatial Resolution &
$\sigma_{xy}$ [cm$^2$]  & $0.1 \times 0.1$   & \multicolumn{2}{c}{$6 \times 6$}  \\
 Energy resolution parameters &
$(S,N,C)$ [\%]    & $(12.2,0,1.4)$ & \multicolumn{2}{c}{$(3,0.2,0.5)$}  \\
\midrule 
\multicolumn{5}{c}{\emph{HCAL}} \\ 
\midrule 
Transverse Spatial Resolution &
$\sigma_{xy}$ [cm$^2$]  & $3 \times 3$   & \multicolumn{2}{c}{$6 \times 6$}  \\
 Energy resolution parameters &
$(S,N,C)$ [\%]    & $(46,0,1.5)$ & \multicolumn{2}{c}{$(30,5,1)$}  \\

\bottomrule
\end{tabular}

\raggedright \small \emph{Note}: For the single-particle relative resolution of the calorimeters, the parametrization 
$\left(\dfrac{\sigma(E)}{E}\right)^2 = \left(\dfrac{S}{\sqrt{E}}\right)^2 + \left(\dfrac{N}{E}\right)^2 + C^2$
is used, where $S,N,C$ are the stochastic, noise and constant terms, representing, respectively the uncertainty due to the intrinsic stochasticity in shower evolution, the readout system's noise, and detector imperfections, shower leakage and other systematic effects~\cite{ParticleDataGroup:2024cfk}.
\end{table*}

\subsection{Flavor tagging algorithm}
To identify the type (``flavor") of the  parton that initiated the formation of the jet we use the \pnetee  \ algorithm~\cite{Bedeschi:2022rnj}; \pnetee\ relies on the \pnet \  algorithm~\cite{Qu:2019gqs} and has been adapted for  \ee \  collision environments. The inputs are processed using a Graph Neural Network (GNN) architecture, where the jet is represented as a graph, with the jet constituents being the nodes of the graph, and relationships between the jet constituents representing the edges of the graph. A series of convolutional operations allows to first explore local patterns and gradually extend to more global ones. After each convolution operation, the node coordinates are dynamically updated based on the learned features, ``grouping" constituents based on their proximity in the input feature space. As laid out in detail in~\cite{Bedeschi:2022rnj}, the \pnetee \ implementation uses a set of 34 input features, covering the relative kinematics of each jet constituent with respect to the jet four-momentum, track-related variables as well as variables related to particle identification. Up to 75 jet constituents are used for each jet. \pnetee \ was chosen as the jet flavor tagging algorithm for this study as a reasonable compromise between computationally lighter architectures and better-performing, yet more compute-intensive models, such as the transformer-based \ParT~\cite{Qu:2022mxj}. For the purpose of our studies, the relative performance differences between detector configurations are more relevant than achieving the highest discrimination power possible.

The output of \pnetee \  is a set of five real numbers $\{p_{j} \},j = q,c,s,b,g$ bounded between 0 and 1, which are the softmax probabilities of a jet belonging to each flavor category. The loss function minimized during the training is the cross-entropy loss:

\begin{equation}
        CE=-\frac{1}{N}{\sum_{i=1}^{N}\sum_{j=1}^{5}{y_{j}^{(i)}\log{p_{j}^{(i)}}}}
    \label{eq:cross_entropy}
\end{equation}

\noindent where $N$ the number of jets in the training dataset, $y_{j}^{(i)}$ a binary number which is one if and only if the $i$-th jet belongs to the $j$-th flavor class and $p_{j}^{(i)}$ the corresponding output probability.

For any jet flavor pair $(i,j)$, the binary discriminant

\begin{equation}
    D_{ij}=\frac{p_i}{p_i+p_j}
    \label{eq:Dij_discr}
\end{equation}

\noindent is used to ascertain the efficiency of tagging flavor $i$ as a function of the mistag rate of background flavor $j$ and build the corresponding receiver operating characteristic (ROC) curve.

\subsection{Simulated Data}
\label{subsec:simulated_data}

The main set of simulated samples consists of Higgstrahlung \ee$ \rightarrow ZH$ events at a center-of-mass (CoM) energy of $\sqrt{s}=250 \ \mathrm{GeV}$, with the $Z$ boson decaying to neutrinos and the Higgs boson decaying hadronically $H\rightarrow j\bar{j}$, $j=q,c,s,b,g$. For each Higgs decay channel, $2\cdot 10^{6}$ events were simulated. The \whiz~\cite{Kilian:2007gr,Moretti:2001zz} generator was used for the hard scatter process, with \py~\cite{Sjostrand:2006za} subsequently used to treat the parton-shower and hadronization. Final-state particles are then reconstructed as PFOs in \delphes, using all of the detector configurations laid out earlier. Finally, within \texttt{FCCAnalyses}~\cite{fccana}, these PFOs are clustered into jets with \fastjet~\cite{Cacciari:2011ma} using the exclusive Durham $k_{T}$ algorithm~\cite{Catani:1991hj} with a fixed number of jets $N_{\mathrm{jets}}=2$ and E-scheme recombination\footnote{Although Durham has been considered the standard jet clustering algorithm for \ee \ collisions, it would be worth evaluating alternatives, such as the generalized $ee-k_{\mathrm{T}}$ algorithm~\cite{Cacciari:2011ma}, and studying the dependence of jet flavor tagging performance on the specific choice of clustering algorithm.}.

To study the CoM energy dependence of our results, the procedure above was also used to obtain $ZHH, Z \rightarrow \nu \bar{\nu}, H \rightarrow j \bar{j}$ events at $\sqrt{s}=550 \ \mathrm{GeV}$. In the latter case, the exclusive jet clustering was done with $N_{\mathrm{jets}}=4$.

The \weaver~\cite{weaver} package was used for the training and the \ranger~\cite{Ranger} \ optimizer was chosen with a starting learning rate of 0.005. The batch size used was 512 and the training was carried out over 35 epochs, with a $70\%-15\%-15\%$ training-validation-testing split. The average accuracy of assigning each jet in the validation set to its correct flavor is used as a metric to select the optimal epoch.

\section{Results}
\label{sec:results}

In this Section, we present the results of the jet flavor tagging studies under the three different detector configurations. 
In \cref{subsec:3_detectors}, we compare the flavor tagging performance for the three different detector concepts introduced in~\cref{subsec:detector_models}. In \cref{subsec:sid_variations}, we study different detector variations for the SiD concept and evaluate their impact on flavor tagging. Finally, in \cref{subsec:energy_dependence}, we assess the robustness of our results with respect to the CoM energy of the samples used.

\subsection{Flavor tagging performance for different detector concepts}
\label{subsec:3_detectors}

\begin{figure*}[ht!]
     \centering
     \begin{subfigure}[b]{0.40\textwidth}
         \centering
         \includegraphics[width=\textwidth]{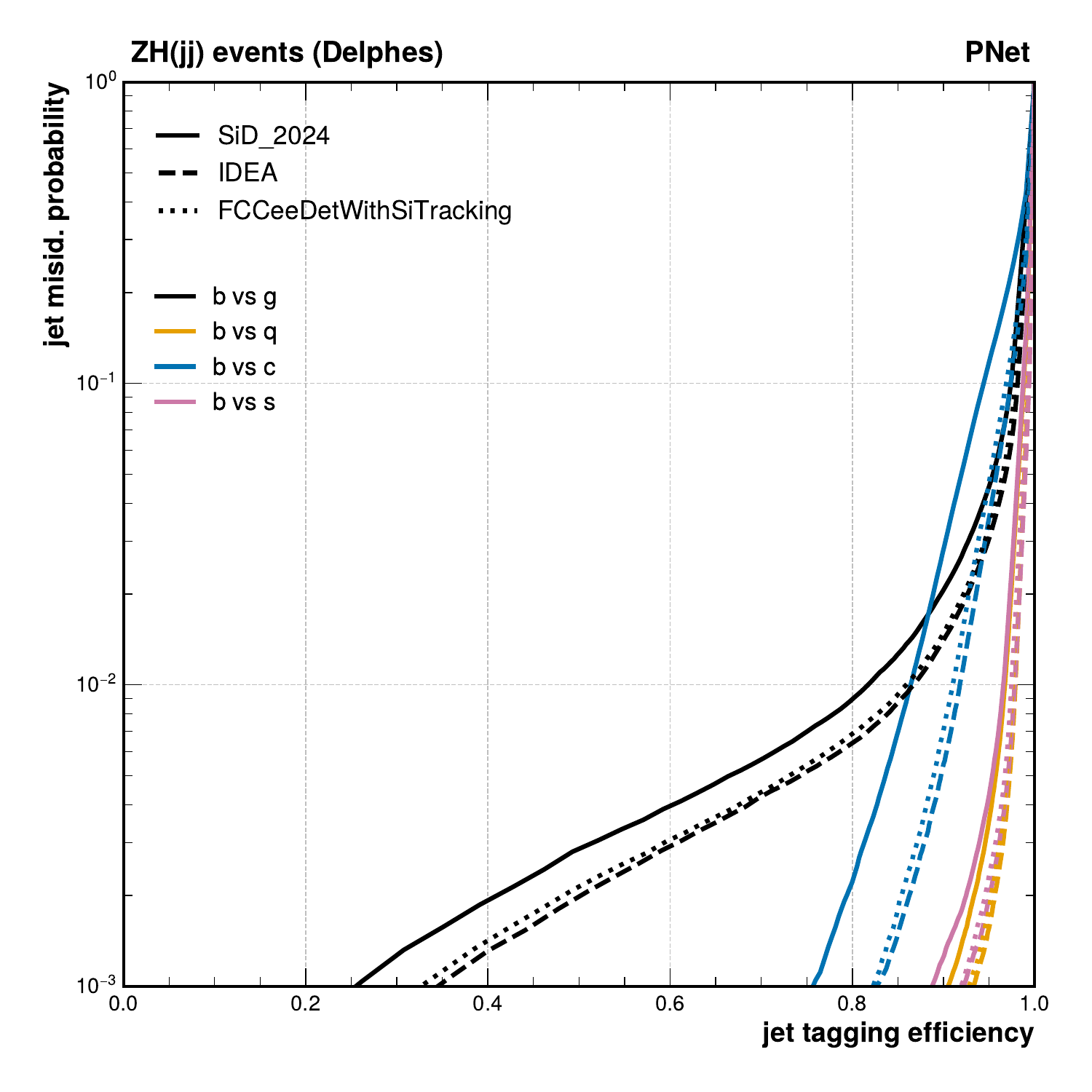}
         \caption{}
         \label{fig:ROC_curves_3_detectors_b_tagging}
     \end{subfigure}
     \begin{subfigure}[b]{0.40\textwidth}
         \centering
         \includegraphics[width=\textwidth]{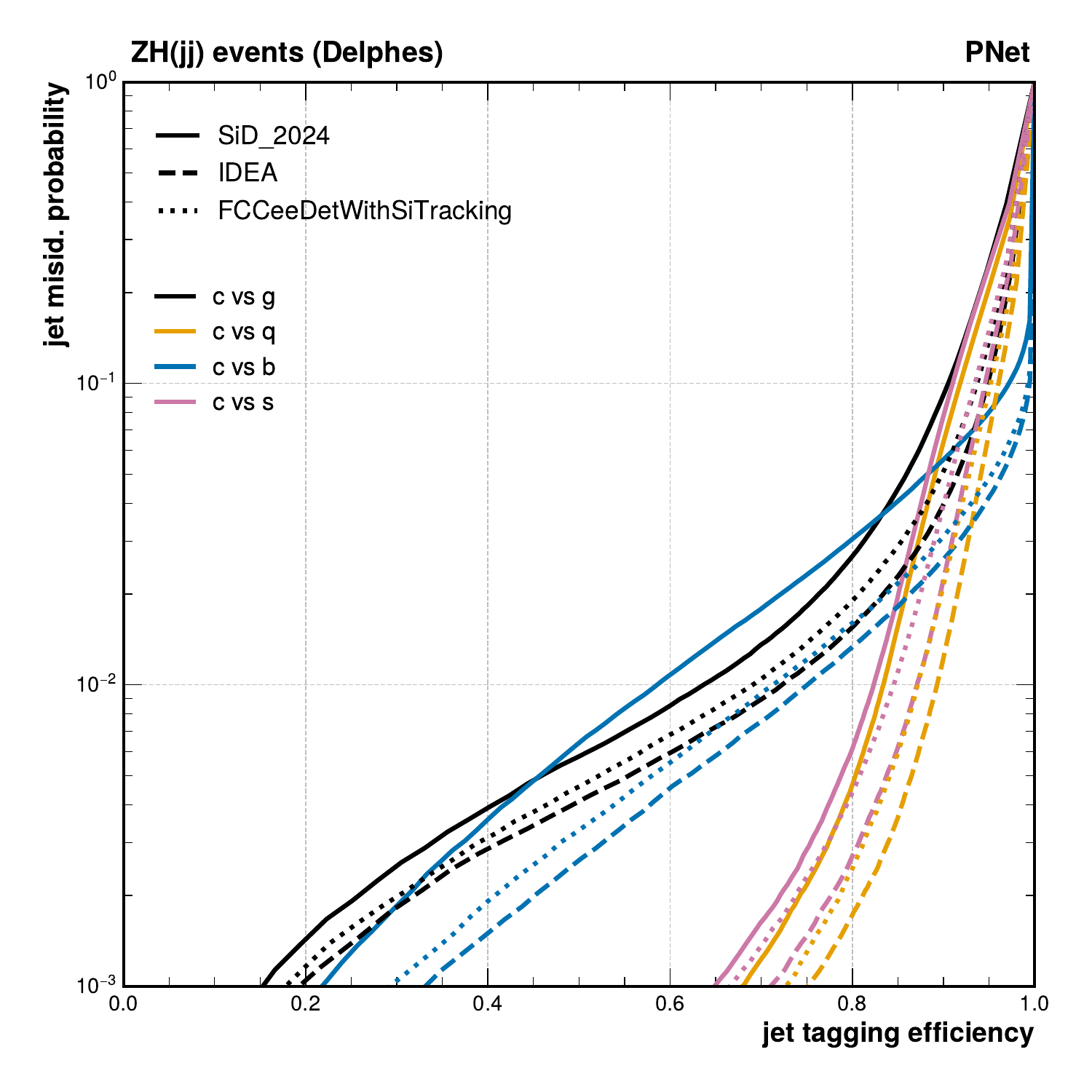}
         \caption{}
         \label{fig:ROC_curves_3_detectors_c_tagging}
     \end{subfigure}
     \begin{subfigure}[b]{0.40\textwidth}
         \centering
         \includegraphics[width=\textwidth]{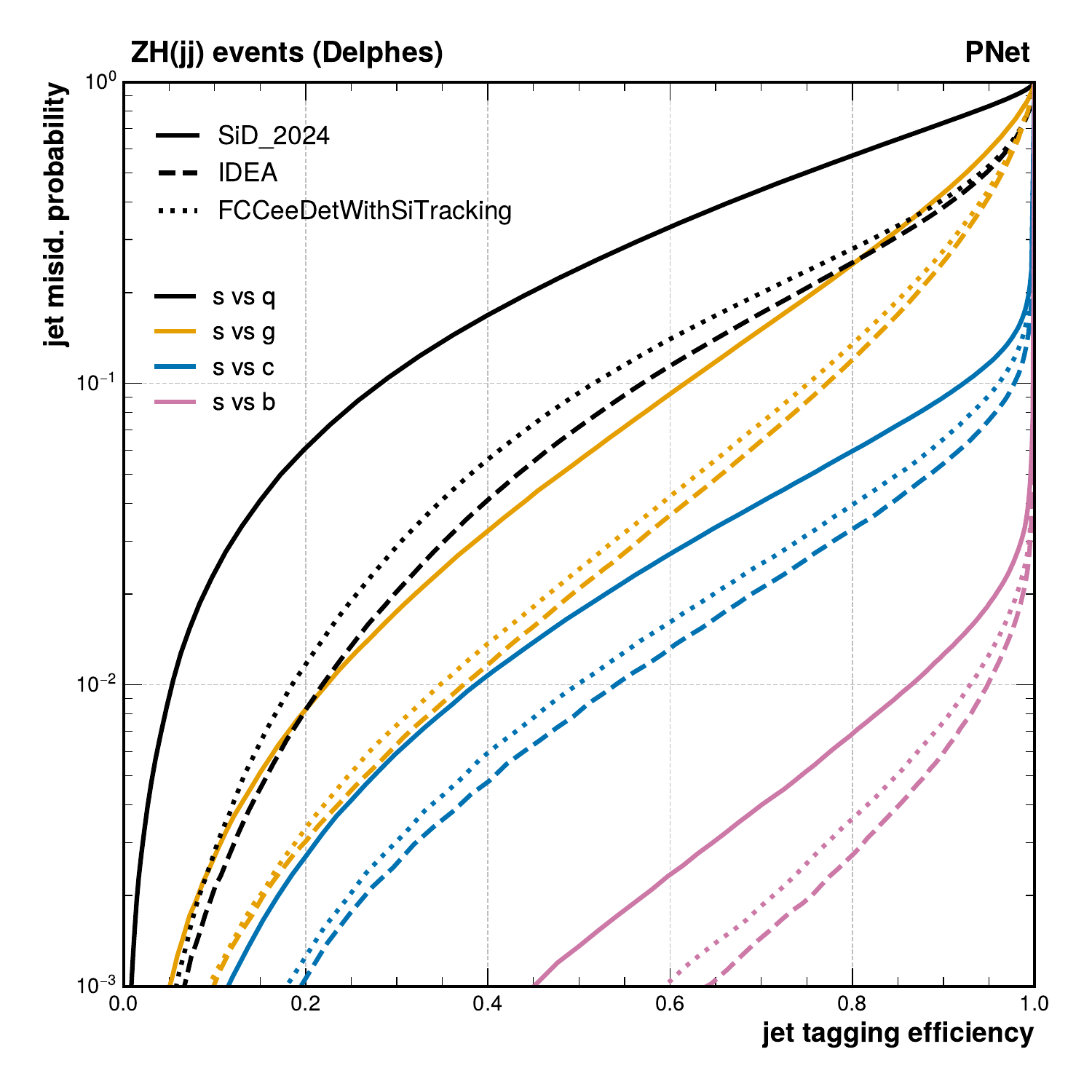}
         \caption{}
         \label{fig:ROC_curves_3_detectors_s_tagging}
     \end{subfigure}
      \begin{subfigure}[b]{0.40\textwidth}
     \centering
     \includegraphics[width=\textwidth]{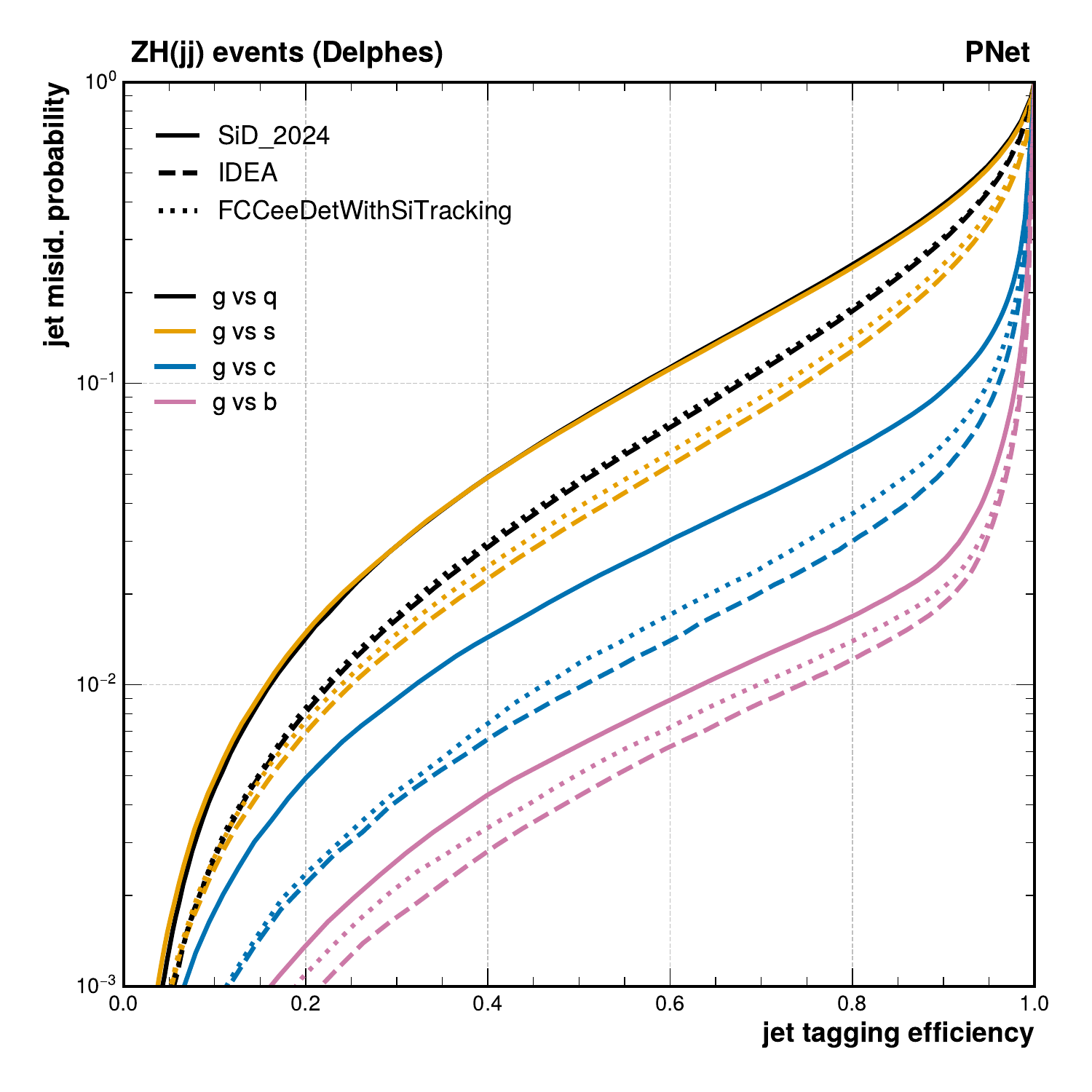}
     \caption{}
     \label{fig:ROC_curves_3_detectors_g_tagging}
        \end{subfigure}
        \caption{ROC curves for the SiD, IDEA, and FCCeeDetWithSiTracking detector concepts where the target task is (a)\emph{b}-tagging, (b)\emph{c}-tagging, (c)\emph{s}-tagging, and (d)\emph{g}-tagging. The horizontal axes indicate the tagging efficiency for the target flavor class, whereas the vertical axes show the probability of mis-identifying a background jet flavor as the target. The different jet flavors considered background
        in each case are indicated in the legends.}
        \label{fig:ROC_curves_3_detectors}
\end{figure*}

Using the analysis framework detailed above, the results obtained for the SiD, IDEA and FCCeeDetWithSiTracking using the $ZH$ Higgstrahlung samples are shown in~\cref{fig:ROC_curves_3_detectors}, where the ROC curves of the mis-identifation probability --- or mistag rate (MR) ---  of background flavor jets has been plotted as a function of the target (or signal) jet flavor efficiency $\epsilon_{S}$. To facilitate the comparison, the mistag rates obtained from \cref{fig:ROC_curves_3_detectors}  for two fixed-signal-efficiency working points (WPs), one at $80\%$ (medium) and one at $90\%$ (loose), are summarized in Table~\ref{tab:detector_comparison_all}. The corresponding Area-Under-Curve (AUC) integrals, defined as 

\begin{equation}
    \mathrm{AUC}^{(B)} =1-\int_{0}^{1}{{\mathrm{MR}^{(B)}}(\epsilon_{S})\mathrm{d}\epsilon_{S}} 
\end{equation}
\noindent where $(B)$ denotes the background jet flavor and $\mathrm{MR}^{(B)}$ the corresponding mistag rate, are also given. AUCs close to unity indicate high discriminating power of the flavor tagger.

Comparing the results for the three detectors in \cref{fig:ROC_curves_3_detectors} and \cref{tab:detector_comparison_all}, we note that the jet tagging performance is overall comparable for the three detector concepts, with overall superior flavor discrimination obtained at IDEA, followed by FCCeeDetWithSiTracking and SiD. Between SiD and the other two detector concepts, most of the differences stem from the lack of PID capabilities for SiD. This is mostly apparent in the case of s-tagging, cf. \cref{fig:ROC_curves_3_detectors_s_tagging}, where the ability to distinguish $s$-hadrons from protons and pions is crucial for tagging $s$-quark initiated jets. Indeed, the mistag rates at both the $80 \%$ and $90 \%$ efficiency for $s$-quarks WPs are $1.4-2.5$ times larger at SiD compared to the other two detectors. A smaller yet non-negligible effect of PID also appears for g-tagging, cf. \cref{fig:ROC_curves_3_detectors_g_tagging}, where the corresponding mistag rate ratio is $1.2-2$. We remind that for both IDEA and FCCeeDetWithSiTracking, TOF measurements with a resolution of 30 ps, as well as cluster counting information assuming the IDEA drift chamber contribute to the detectors' PID capabilities, with the latter providing most of the gain~\cite{Bedeschi:2022rnj}.  The remaining discrepancies between FCCeeDetWithSiTracking and IDEA come from the improved vertex and tracking system of IDEA.

Regardless of the specific detector choice, the predicted flavor tagging performance is overall pristine, reaching mistag rates at the percent and sub-percent levels for $b$- and $c$- tagging, while preserving high signal purity, and enabling the identification of $s$- and $g$-jets with sufficient accuracy. This represents an order of magnitude improvement with respect to the latest results for the CMS~\cite{CMS-DP-2024-066} and ATLAS~\cite{ATLAS:2022rkn} experiments at the LHC, owing to the stringent precision requirements  these future detector concepts have been developed to meet, and the extremely clean experimental environment at an \ee  \ collider, which doesn't suffer from the large amounts of pile-up and underlying event particles contributing to the final state events at the LHC\footnote{It is worth noting that beam-induced background (BIB) particles are also present at \ee \  colliders, mostly in the form of Beamstrahlung \ee \ pairs and low-transverse-momentum photoproduced hadrons. Such backgrounds were not taken into account in the studies presented here but are expected to have a small impact on flavor tagging performance.}.

\begin{table*}[htbp]
    \centering
    \begin{tabular}{c@{\hspace{2em}}!{\vline width 3\arrayrulewidth}@{\hspace{2em}}c}  

%

        \begin{tabular}{cc|cccc}
            \multicolumn{6}{c}{\Large\textbf{b-tagging}} \\[1em]
            \multicolumn{2}{c|}{} & \multicolumn{4}{c}{\textbf{Mistag rates}} \\ 
            \hline 
            WP & Eff (b) & \emph{b} vs \emph{g} & \emph{b} vs \emph{q} & \emph{b} vs \emph{c} & \emph{b} vs \emph{s} \\
            \hline
            & & \textcolor{CBBlue}{2.1\%} & \textcolor{CBBlue}{$<$0.1 \%} & \textcolor{CBBlue}{2.9\%} & \textcolor{CBBlue}{0.13 \%} \\
            Loose & 90\% &  \textcolor{CBOrange}{1.5\%} & \textcolor{CBOrange}{$<$0.1 \%} & \textcolor{CBOrange}{ 0.72 \%} & \textcolor{CBOrange}{$<$0.1 \%} \\
            & & \textcolor{CBPurple}{1.4\%} & \textcolor{CBPurple}{$<$0.1 \%} & \textcolor{CBPurple}{0.55 \%} & \textcolor{CBPurple}{$<$0.1 \%} \\
            \hline
            & & \textcolor{CBBlue}{0.88\%} & \textcolor{CBBlue}{$<$0.1 \%} & \textcolor{CBBlue}{0.22\%} & \textcolor{CBBlue}{$<$0.1 \%} \\
            Medium & 80\% & \textcolor{CBOrange}{0.69\%} & \textcolor{CBOrange}{$<$0.1 \%} & \textcolor{CBOrange}{$< $0.1 \%} & \textcolor{CBOrange}{$<$0.1 \%} \\
            & & \textcolor{CBPurple}{0.64\%} & \textcolor{CBPurple}{$<$0.1 \%} & \textcolor{CBPurple}{$<$0.1 \%} & \textcolor{CBPurple}{$<$0.1 \%} \\
            \hline\hline
            \multicolumn{2}{c|}{\textbf{Area Under}} & \textcolor{CBBlue}{0.9863} & \textcolor{CBBlue}{1.0000} & \textcolor{CBBlue}{0.9821} & \textcolor{CBBlue}{0.9998} \\
            \multicolumn{2}{c|}{\textbf{Curve}} & \textcolor{CBPurple}{0.9901} & \textcolor{CBPurple}{1.0000} & \textcolor{CBPurple}{0.9909} & \textcolor{CBPurple}{1.0000} \\
            \multicolumn{2}{c|}{\textbf{(AUC)}}  & \textcolor{CBOrange}{0.9910} & \textcolor{CBOrange}{1.0000} & \textcolor{CBOrange}{0.9931} & \textcolor{CBOrange}{1.0000} \\
            \hline
        \end{tabular}
        & 
        \begin{tabular}{cc|cccc}
            \multicolumn{6}{c}{\Large\textbf{c-tagging}} \\[1em]
            \multicolumn{2}{c|}{} & \multicolumn{4}{c}{\textbf{Mistag rates}} \\ 
            \hline 
            WP & Eff (c) & \emph{c} vs \emph{g} & \emph{c} vs \emph{q} & \emph{c} vs \emph{b} & \emph{c} vs \emph{s} \\
            \hline
            & & \textcolor{CBBlue}{9.3\%} & \textcolor{CBBlue}{6.4\%} & \textcolor{CBBlue}{5.5\%} & \textcolor{CBBlue}{7.6\%} \\
            Loose & 90\% & \textcolor{CBOrange}{5.2\%} & \textcolor{CBOrange}{2.2\%} & \textcolor{CBOrange}{3.2\%} & \textcolor{CBOrange}{4.1\%} \\
            & & \textcolor{CBPurple}{4.0\%} & \textcolor{CBPurple}{1.2\%} & \textcolor{CBPurple}{2.6\%} & \textcolor{CBPurple}{2.2\%} \\
            \hline
            & & \textcolor{CBBlue}{2.6\%} & \textcolor{CBBlue}{4.6\%} & \textcolor{CBBlue}{3.0\%} & \textcolor{CBBlue}{0.62\%} \\
            Medium & 80\% & \textcolor{CBOrange}{1.9\%} & \textcolor{CBOrange}{0.24\%} & \textcolor{CBOrange}{1.6\%} & \textcolor{CBOrange}{0.44\%} \\
            & & \textcolor{CBPurple}{1.5\%} & \textcolor{CBPurple}{0.17\%} & \textcolor{CBPurple}{1.3\%} & \textcolor{CBPurple}{0.27\%} \\
            \hline\hline
            \multicolumn{2}{c|}{\textbf{Area Under}} & \textcolor{CBBlue}{0.9592} & \textcolor{CBBlue}{0.9729} & \textcolor{CBBlue}{0.9982} & \textcolor{CBBlue}{0.9681} \\
             \multicolumn{2}{c|}{\textbf{Curve}} & \textcolor{CBOrange}{0.9782} & \textcolor{CBOrange}{0.9920} & \textcolor{CBOrange}{1.0000} & \textcolor{CBOrange}{0.9822} \\
        \multicolumn{2}{c|}{\textbf{(AUC)}} & \textcolor{CBPurple}{0.9805} & \textcolor{CBPurple}{0.9929} & \textcolor{CBPurple}{1.0000} & \textcolor{CBPurple}{0.9858} \\
            \hline
        \end{tabular}
        \\ \\ \\   \Xhline{3\arrayrulewidth} \\ \\ 
        \begin{tabular}{cc|cccc}
            \multicolumn{6}{c}{\Large\textbf{s-tagging}} \\[1em]
            \multicolumn{2}{c|}{} & \multicolumn{4}{c}{\textbf{Mistag rates}} \\ 
            \hline 
            WP & Eff (s) & \emph{s} vs \emph{q} & \emph{s} vs \emph{g} & \emph{s} vs \emph{c} & \emph{s} vs \emph{b} \\
            \hline
            & & \textcolor{CBBlue}{70\%} & \textcolor{CBBlue}{45\%} & \textcolor{CBBlue}{9.1\%} & \textcolor{CBBlue}{1.3\%} \\
            Loose & 90\% & \textcolor{CBOrange}{41\%} & \textcolor{CBOrange}{29\%} & \textcolor{CBOrange}{6.4\%} & \textcolor{CBOrange}{0.76\%} \\
            & & \textcolor{CBPurple}{38\%} & \textcolor{CBPurple}{26\%} & \textcolor{CBPurple}{5.4\%} & \textcolor{CBPurple}{0.59\%} \\
            \hline
            & & \textcolor{CBBlue}{54\%} & \textcolor{CBBlue}{24\%} & \textcolor{CBBlue}{6.0\%} & \textcolor{CBBlue}{0.68\%} \\
            Medium & 80\% & \textcolor{CBOrange}{28\%} & \textcolor{CBOrange}{13\%} & \textcolor{CBOrange}{4.0\%} & \textcolor{CBOrange}{0.36\%} \\
            & & \textcolor{CBPurple}{25\%} & \textcolor{CBPurple}{12\%} & \textcolor{CBPurple}{3.3\%} & \textcolor{CBPurple}{0.27\%} \\
            \hline\hline
            \multicolumn{2}{c|}{\textbf{Area Under}} & \textcolor{CBBlue}{0.6866} & \textcolor{CBBlue}{0.8575} & \textcolor{CBBlue}{0.9712} & \textcolor{CBBlue}{1.0000} \\
            \multicolumn{2}{c|}{\textbf{Curve}} & \textcolor{CBOrange}{0.8439} & \textcolor{CBOrange}{0.9091} & \textcolor{CBOrange}{0.9813} & \textcolor{CBOrange}{1.0000} \\
        \multicolumn{2}{c|}{\textbf{(AUC)}} & \textcolor{CBPurple}{0.8598} & \textcolor{CBPurple}{0.9160} & \textcolor{CBPurple}{0.9866} & \textcolor{CBPurple}{1.0000} \\
            \hline
        \end{tabular}
        &
        \begin{tabular}{cc|cccc}
            \multicolumn{6}{c}{\Large\textbf{g-tagging}} \\[1em]
            \multicolumn{2}{c|}{} & \multicolumn{4}{c}{\textbf{Mistag rates}} \\ 
            \hline 
            WP & Eff (g) & \emph{g} vs \emph{q} & \emph{g} vs \emph{s} & \emph{g} vs \emph{c} & \emph{g} vs \emph{b} \\
            \hline
            & & \textcolor{CBBlue}{41\%} & \textcolor{CBBlue}{39\%} & \textcolor{CBBlue}{9.5\%} & \textcolor{CBBlue}{2.6\%} \\
            Loose & 90\% & \textcolor{CBOrange}{31\%} & \textcolor{CBOrange}{24\%} & \textcolor{CBOrange}{6.4\%} & \textcolor{CBOrange}{2.1\%} \\
            & & \textcolor{CBPurple}{31\%} & \textcolor{CBPurple}{23\%} & \textcolor{CBPurple}{5.1\%} & \textcolor{CBPurple}{1.9\%} \\
            \hline
            & & \textcolor{CBBlue}{26\%} & \textcolor{CBBlue}{25\%} & \textcolor{CBBlue}{6.1\%} & \textcolor{CBBlue}{1.7\%} \\
            Medium & 80\% & \textcolor{CBOrange}{18\%} & \textcolor{CBOrange}{14\%} & \textcolor{CBOrange}{3.7\%} & \textcolor{CBOrange}{1.4\%} \\
            & & \textcolor{CBPurple}{17\%} & \textcolor{CBPurple}{13\%} & \textcolor{CBPurple}{3.0\%} & \textcolor{CBPurple}{1.2\%} \\
            \hline\hline
            \multicolumn{2}{c|}{\textbf{Area Under}} & \textcolor{CBBlue}{0.8551} & \textcolor{CBBlue}{0.8576} & \textcolor{CBBlue}{0.9596} & \textcolor{CBBlue}{0.9895} \\
            \multicolumn{2}{c|}{\textbf{Curve}} & \textcolor{CBOrange}{0.8904} & \textcolor{CBOrange}{0.9092} & \textcolor{CBOrange}{0.9730} & \textcolor{CBOrange}{0.9920} \\
        \multicolumn{2}{c|}{\textbf{(AUC)}} & \textcolor{CBPurple}{0.8927} & \textcolor{CBPurple}{0.9163} & \textcolor{CBPurple}{0.9781} & \textcolor{CBPurple}{0.9942} \\
            \hline
        \end{tabular}
    \end{tabular}
    \caption{Comparison of \pnetee \  flavor tagging performance for different detector concepts: SiD (in \textcolor{CBBlue}{blue}), FCCeeDetWithSiTracking (in \textcolor{CBOrange}{orange}) and IDEA (in \textcolor{CBPurple}{magenta}). For each detector and target jet flavor, the background mistag rates for the $80\%$ and $90\%$ fixed signal efficiency working points, as well as the Area-Under-Curve values of the corresponding ROC curves of \cref{fig:ROC_curves_3_detectors}, are given.}
    \label{tab:detector_comparison_all}
\end{table*}

\subsection{SiD detector variations}
\label{subsec:sid_variations}

In addition to the studies on the dependence of the flavor tagging performance on detector configuration, which generally have sizeable differences in their design and performance, it is worth evaluating how finer modifications in detector design translate to changes in jet tagging performance. Such studies are important for informing potential directions for detector optimization, given that the finalized designs of future detectors for \ee \ colliders lie several years in the future.

Similar detector variations studies have recently been performed for IDEA, but are mostly focused on examining variations in the vertex detector~\cite{Bedeschi:2022rnj}. In this study, we use the SiD concept as a benchmark and evaluate the impact of modifications not only on the vertex system but also on the energy and spatial resolution of both the ECAL and HCAL.

\begin{figure*}[htbp]
    \centering
    \includegraphics[width=\linewidth]{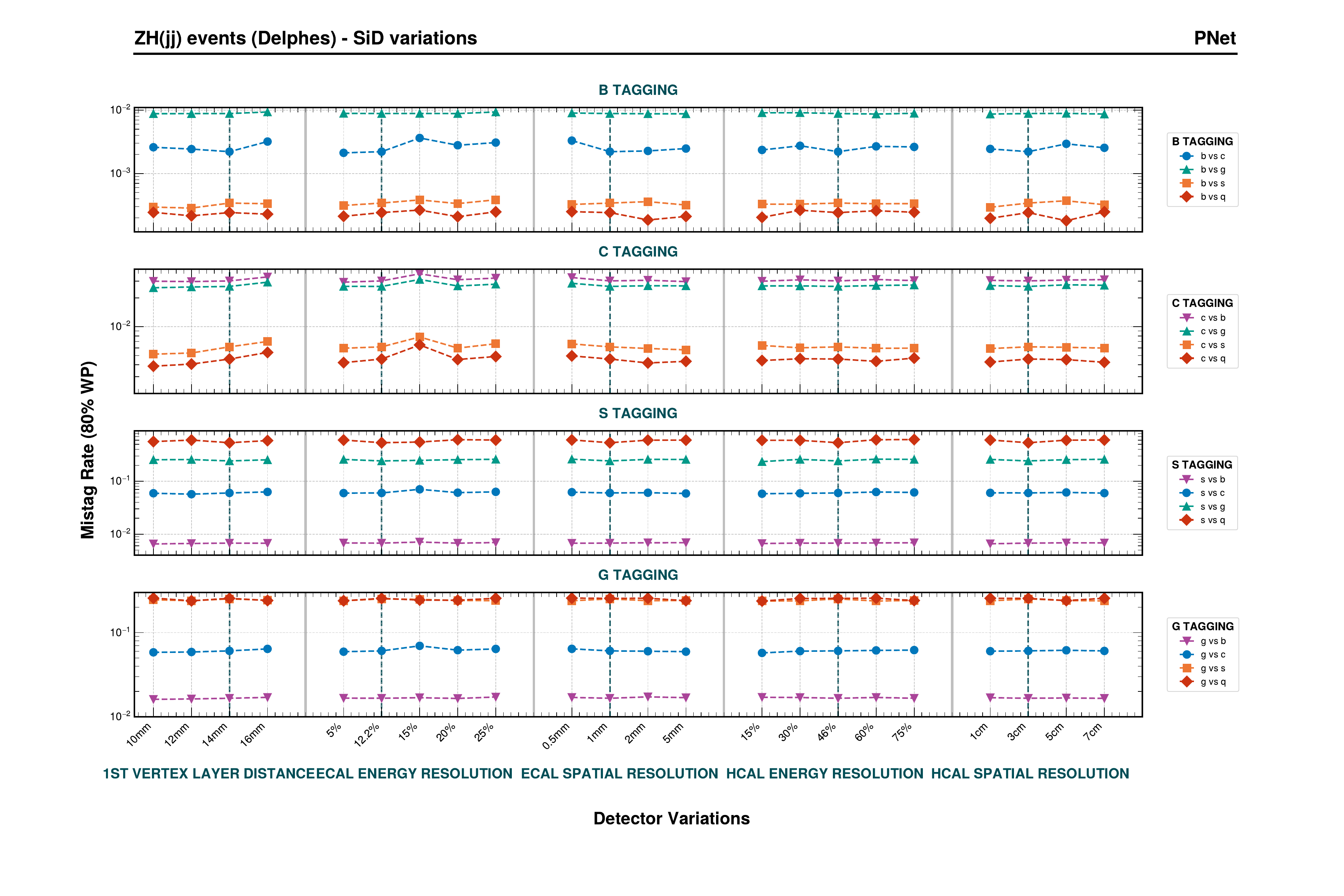}
    \caption{Mistag rates for the $80\%$ signal efficiency working point for several variations of the SiD baseline design. The four panes correspond to different target jet flavors. For clarity, thicker vertical lines are used to indicate the baseline SiD scenario in each variation case. }
    \label{fig:SiD_variations_mistag_rates}
\end{figure*}

Specifically, for each of the following parameters: radial distance of the first vertex barrel layer from the IP, stochastic term $S$ in the energy resolution parametrization of the ECAL and HCAL (see~\cref{tab:detector_specs}) and transverse spatial resolution of the ECAL  and HCAL, we select a few typical values, representing both improvement as well as degradation with respect to the nominal ones in~\cref{tab:detector_specs}, and repeat the training procedure. We then quantify the jet flavor tagging performance by estimating the mistag rate at the $80\%$ signal efficiency WP. The results of this study are summarized in~\cref{fig:SiD_variations_mistag_rates}.

Jet tagging performance shows relatively stable performance across all tagging categories with respect to the position of the first barrel layer of the vertex detector. This suggests that, within the studied range of $10-16 \ \mathrm{mm}$, the detector's tracking capabilities remain robust. A slight degradation is observed in $b$- and $c$-tagging performance as the distance increases, which is anticipated as larger distances of the vertex detector from the IP make the reconstruction of additional vertices from  $B$- and $D$-hadron decays more challenging.

The ECAL energy and spatial resolution variation shows minimal impact on flavor tagging performance across all categories. This is expected since flavor tagging primarily relies on tracking information and vertex reconstruction rather than calorimeter energy measurements and cluster granularity. While slight variations are observed at poorer resolutions, particularly in the ECAL case where degraded Particle-Flow performance might affect the separation of nearby energy deposits in dense jets, the overall effect remains small.  The stability of mistag rates across different ECAL energy resolution values demonstrates the potency of the Particle-Flow approach, which effectively separates the charged and neutral components of jets, making the flavor tagging robust against calorimeter resolution variations.

Similar to the ECAL case, variations in the energy and spatial resolution of the HCAL show minimal influence on flavor tagging performance, with the mistag rates remaining stable even with significant degradations of the resolution from $15\%$ to $75\%$. The robustness against HCAL resolution variations is particularly important for cost optimization, suggesting that HCAL requirements could potentially be relaxed without significantly compromising flavor tagging capabilities. We do note, however, that other key physics benchmarks require excellent calorimeter performance. Precise ECAL resolution is crucial for $H\rightarrow \gamma \gamma $ measurements, electron identification, photon reconstruction in radiative decays, and $\pi^{0}$ reconstruction in tau decays. The HCAL performance is essential for precise $W/Z$ boson separation in hadronic decays, Higgs branching ratio measurements in multi-jet final states (particularly in $H\rightarrow WW^{*}/ZZ^{*} \rightarrow 4j$ channels), and the reconstruction of hadronic tau decays. Additionally, searches for beyond Standard Model physics involving highly boosted objects or long-lived particles decaying in the calorimeter system necessitate maintaining good energy resolution in both calorimeter systems. These requirements must be carefully balanced against potential cost savings in the detector optimization process.

\subsection{Center-of-mass energy dependence}
\label{subsec:energy_dependence}

All future \ee \  machines have been developed with a staged approach in mind, with physics runs at multiple CoM energy.  In particular, circular colliders, such as the FCC-ee~\cite{FCC:2018evy} and CEPC~\cite{CEPCStudyGroup:2018ghi}, have proposed runs at the \emph{Z}-pole, \emph{WW}-threshold, \emph{ZH}-region and $t\bar{t}$ threshold at CoM energies of $\sqrt{s} \sim 90,160,240$ and $360 \ \mathrm{GeV}$, respectively. Linear colliders, such as ILC~\cite{ILC:2013jhg} and C$^3$~\cite{Vernieri_2023}, have proposed runs at $\sqrt{s}=250, 500-550 \ \mathrm{GeV}$, with higher energies reachable with suitable upgrades. In either of these cases, a jet-flavor tagger trained at a specific CoM energy will later have to be applied at a different one. At a circular collider, the large statistics ($\mathcal{O}(10^{12})$) of \emph{Z}-bosons expected to be produced, with the majority of them decaying hadronically, provides an abundant sample for in-situ calibration of the flavor tagging algorithm. The calibrated tagger can then be applied at subsequent higher energy runs. Similarly, at a linear collider, a tagger trained and calibrated on $ZH$ events at $\sqrt{s}=250 \ \mathrm{GeV}$ could then be applied at the $500-550 \ \mathrm{GeV}$ run.

\begin{figure*}[ht!]
     \centering
     \begin{subfigure}[b]{0.40\textwidth}
         \centering
         \includegraphics[width=\textwidth]{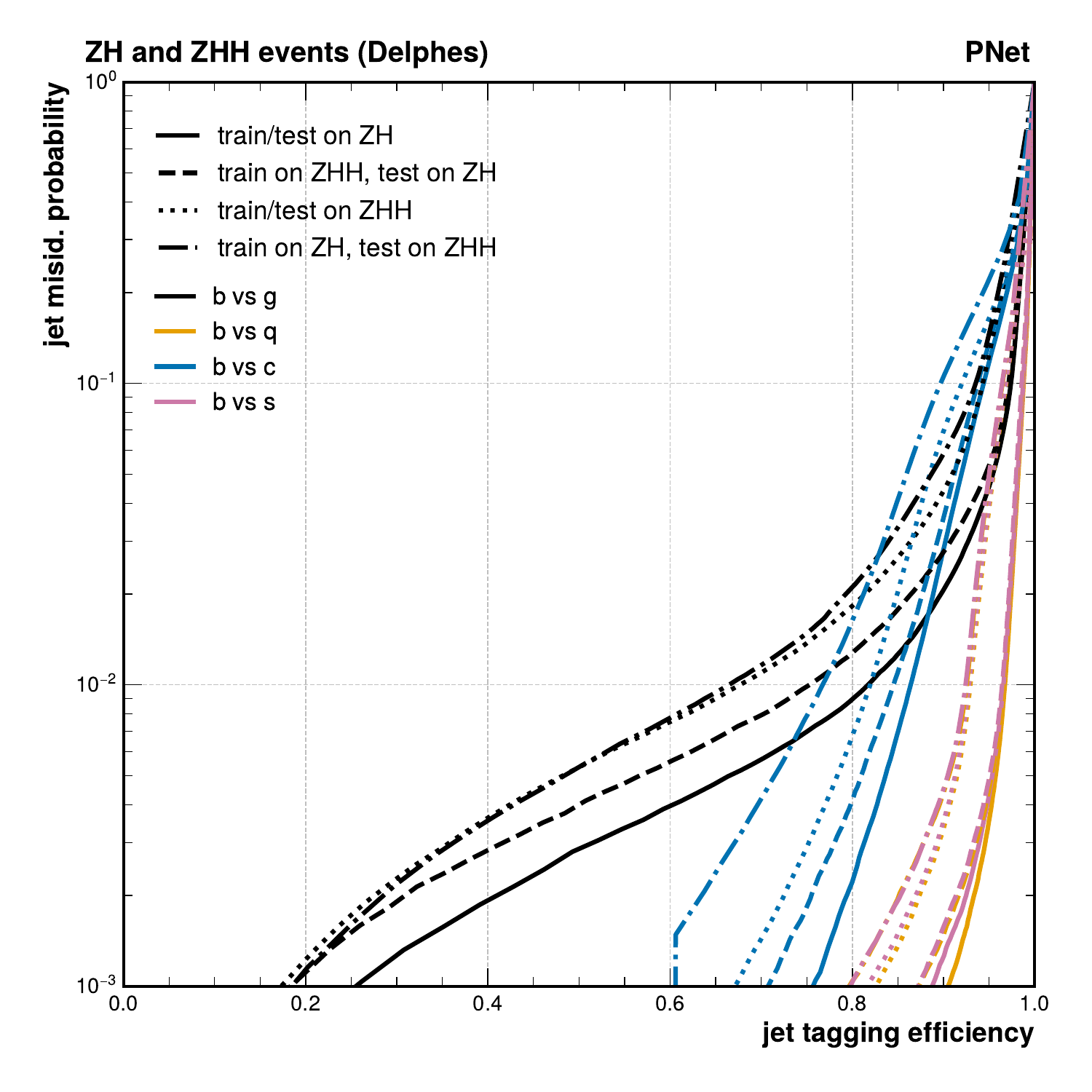}
         \caption{}
         \label{fig:ROC_curves_ZH_vs_ZHH_b_tagging}
     \end{subfigure}
     \begin{subfigure}[b]{0.40\textwidth}
         \centering
         \includegraphics[width=\textwidth]{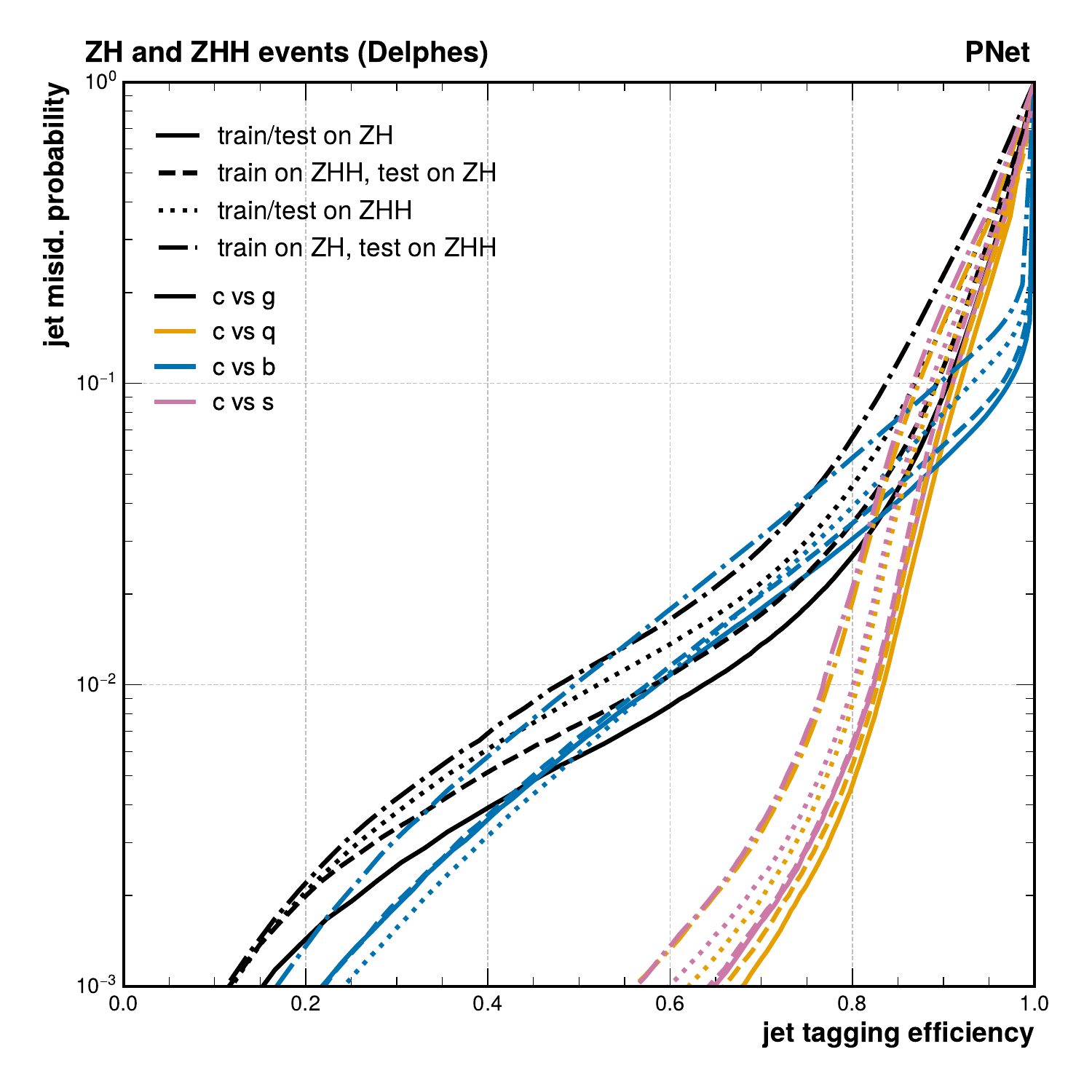}
         \caption{}
         \label{fig:ROC_curves_ZH_vs_ZHH_c_tagging}
     \end{subfigure}
     \begin{subfigure}[b]{0.40\textwidth}
         \centering
         \includegraphics[width=\textwidth]{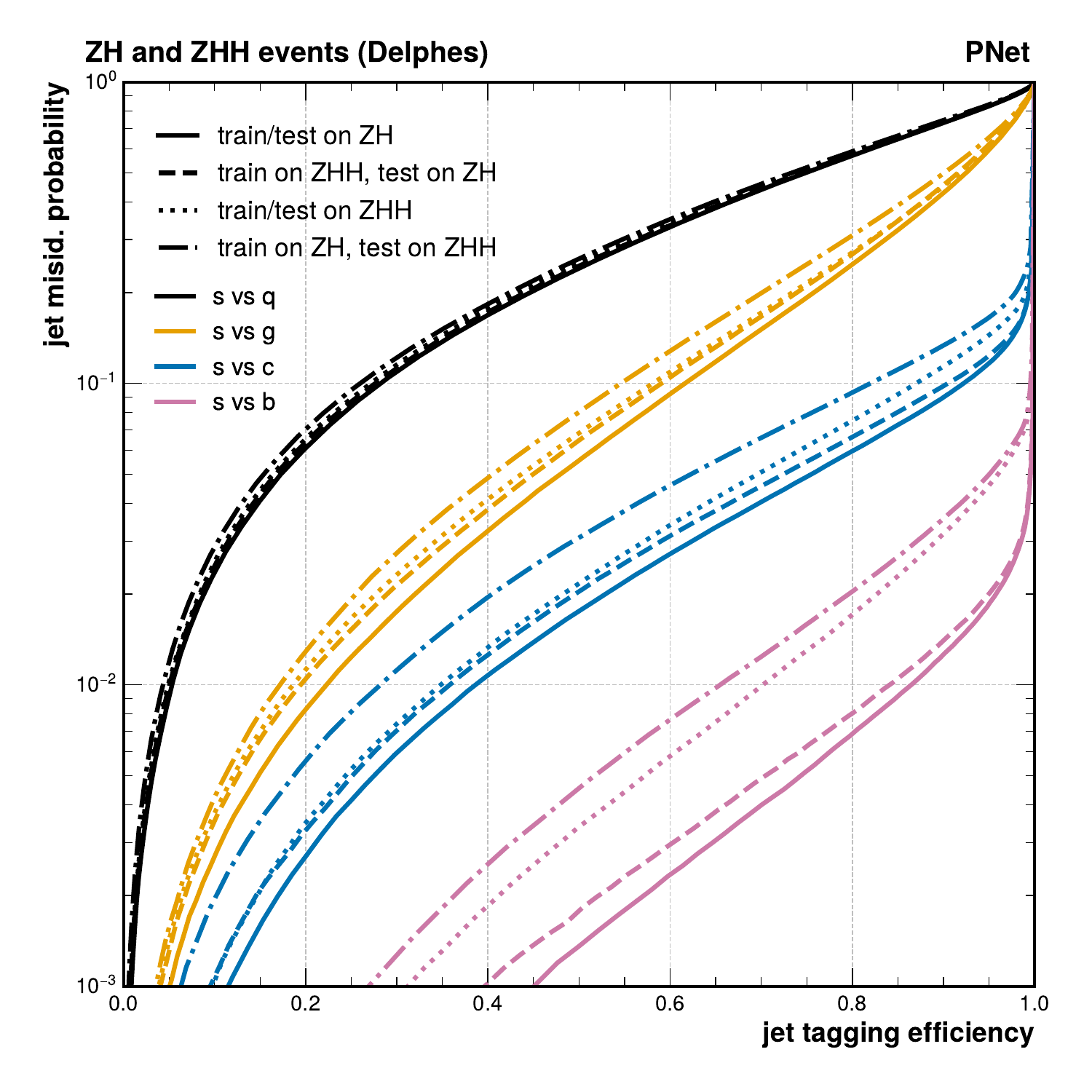}
         \caption{}
         \label{fig:RROC_curves_ZH_vs_ZHH_s_tagging}
     \end{subfigure}
      \begin{subfigure}[b]{0.40\textwidth}
     \centering
     \includegraphics[width=\textwidth]{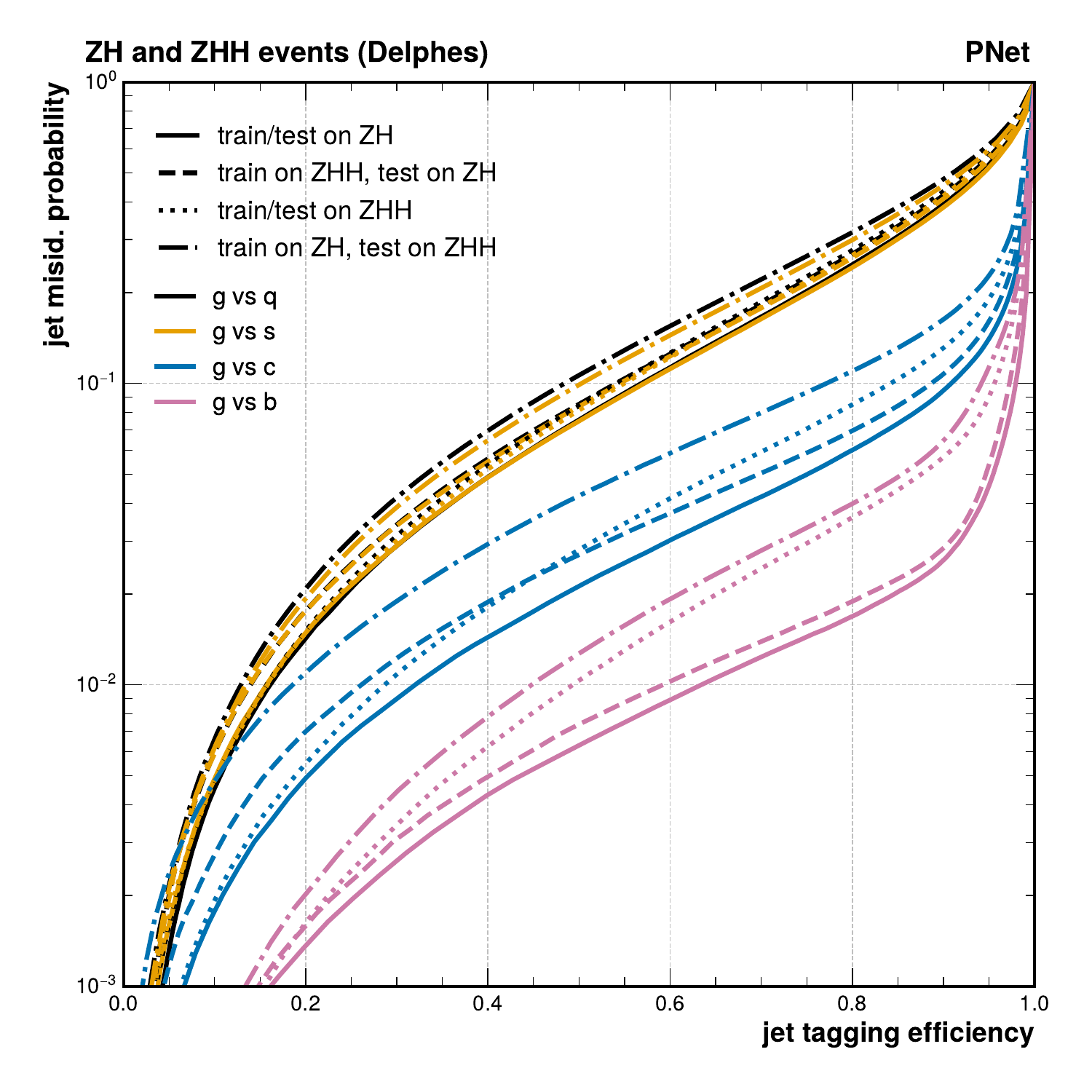}
     \caption{}
     \label{fig:ROC_curves_ZH_vs_ZHH_g_tagging}
        \end{subfigure}
        \caption{ROC curves for the SiD detector concept where the target task is (a)\emph{b}-tagging, (b)\emph{c}-tagging, (c)\emph{s}-tagging, and (d)\emph{g}-tagging and the training and/or inference have been performed on $ZH$ and $ZHH$ samples. The horizontal axes indicate the tagging efficiency for the target flavor class, whereas the vertical axes show the probability of mis-identifying a background jet flavor as the target. The different jet flavors considered background
        in each case are indicated in the legends.}
        \label{fig:ROC_curves_ZH_vs_ZHH}
\end{figure*}

It is therefore an interesting exercise to evaluate how the extracted jet flavor tagging performance depends on the CoM energy of the samples used in the training. With this in mind, as detailed in~\cref{subsec:simulated_data}, we repeat the training for the SiD concept at the additional CoM energy of 550 GeV, using $ZHH$ double-Higgsstralung samples. For each of the two CoM energy scenarios, we test the tagger on samples of either energy, thus effectively simulating how a tagger trained at a specific CoM energy would perform at another one. The results, in the form of the corresponding ROC curves, are presented in \cref{fig:ROC_curves_ZH_vs_ZHH}.

We observe that, among the four train-test combinations shown in \cref{fig:ROC_curves_ZH_vs_ZHH},  there are noticeable differences in the flavor tagging performance, mostly for $b$- and $c$-jets, with up to an order of magnitude different mistag rates for the same signal efficiency. These discrepancies are potentially due to differences in the jet momentum spectrum between the $ZH$ and $ZHH$ final states, with jets in the former case having a more constrained phase-space. We anticipate that a training strategy with matched, reweighted momentum distributions could relieve some of the observed tension.

\section{Conclusions and Future Work}
\label{sec:conclusions}

In this study, we have presented a systematic evaluation of jet flavor tagging performance across different detector concepts for future \ee \  colliders. Using a unified analysis framework relying on \delphes \ fast simulation and the \pnetee \ tagger, we assessed, on an equal footing, the jet flavor discrimination capabilities for different detector concepts while exploring the impact of various detector parameters.    


The studied detector concepts demonstrate significant improvements in flavor tagging performance compared to similar algorithms at the LHC. Among the detector designs, IDEA and FCCeeDetWithSiTracking show superior discrimination power, with their dedicated PID capabilities through time-of-flight and cluster counting measurements emerging as a key differentiating factor. This is particularly evident in $s$-tagging performance, where these detectors achieve up to 2.5 times lower mistag rates compared to SiD.

Furthermore, a systematic investigation of detector parameter variations for the updated SiD concept was pursued for the first time, exhibiting the robustness of flavor tagging performance against changes in calorimeter energy and spatial resolution, while showing only modest degradation in heavy flavor tagging performance when increasing the radius of the first vertex barrel layer up to $16 \ \mathrm{mm}$.  Although this resilience suggest potential opportunities for cost optimization in the calorimeter systems without significantly compromising flavor tagging capabilities, we stress the importance of maintaining excellent calorimeter performance for other key physics benchmarks

The framework developed in this study can be readily extended to evaluate additional detector variations and their impact on flavor tagging performance. Future work could include studying the impact of different magnetic field strengths, investigating alternative detector technologies, and exploring the effects of beam-induced backgrounds. Furthermore, the observation of center-of-mass energy dependence in tagging performance suggests the need for developing robust training strategies that maintain consistent performance across different collision energies.

While our study provides insights for detector optimization, several limitations and areas for future work should be noted. First, the results presented here rely on fast simulation, and validation against full, \geant-based detector simulation will be essential to confirm the observed trends and quantify potential simulation artifacts. Additionally, our parameter variations were conducted as one-dimensional scans, keeping other parameters fixed at their nominal values. A natural extension would be to explore the multi-dimensional parameter space through systematic multi-objective optimization, while simultaneously taking into account various physics benchmarks beyond flavor tagging.

\section{Acknowledgements}
The authors express their gratitude to Michele Selvaggi, Jim Brau and Jan Strube for their inputs and insightful discussions, which have significantly contributed to this study.

The work of D.N. and C.V. is supported by the US Department of Energy under contract DE–AC02–76SF00515.

This work used the resources of the SLAC Shared Science Data Facility (S3DF) at SLAC National Accelerator Laboratory.  SLAC is operated by Stanford University for the U.S. Department of Energy’s Office of Science.



\section{Code and Data Availability}

The code used to obtain the results of this study can be found at \url{https://github.com/dntounis/jet-flavor-tagging-ee}. The corresponding data files are available upon request from the authors.

\bibliography{bibliography} 

\end{document}